\pdfoutput=1  
\documentclass[twocolumn,noarxiv]{quantumarticle}

\usepackage[utf8]{inputenc}
\usepackage{amsmath,amssymb,amsthm}
\usepackage{booktabs}
\newcommand{\rr}{\raggedright\let\\\tabularnewline}
\usepackage{multirow}
\usepackage{enumitem}
\usepackage{graphicx}
\usepackage{xcolor}
\usepackage[colorlinks=true,allcolors=blue]{hyperref}
\usepackage{orcidlink}
\usepackage[round,authoryear]{natbib}
\graphicspath{{figs/}}

\newcommand{\pillar}[1]{\textsc{#1}}

\begin{document}

\title{What kind of quantum advantage is this?}

\author{Christophe Pere\,\orcidlink{0000-0002-8902-787X}}
\affiliation{\'Ecole de Technologie Sup\'erieure, Universit\'e du Qu\'ebec, 1100 Notre-Dame St W, Montreal, Quebec H3C 1K3, Canada}
\affiliation{Universit\'e Laval, 1045, avenue de la M\'edecine, Qu\'ebec, Quebec G1V 0A6, Canada}

\date{\today}

\begin{abstract}
Quantum advantage encompasses fundamentally different claims, from resource reductions to exponential separations in sample complexity. Evaluating these claims requires distinguishing how an algorithm achieves an advantage from what is necessary for the separation and how strongly it is supported against classical alternatives. We propose a multi-label taxonomy that makes these distinctions explicit, classifying structural, interference, and coherent mechanisms while separately recording their necessity and the evidence for the classical comparison. Coherent mechanisms are further divided into memory and output, reflecting distinct restrictions on quantum information processing. A two-part entanglement-breaking test distinguishes loss of information from loss of algorithmic efficiency. The hidden subgroup problem illustrates the distinction: among the cases examined, cross-copy coherence is information-theoretically necessary only for the symmetric group. We present this as a diagnostic methodology rather than a theory of quantum advantage: for a new claim, it identifies the relevant adversary, separates the mechanisms a protocol uses from those shown to be necessary, and records the strength of the evidence and the quality of the advantage separately.
\end{abstract}

\maketitle

\section{Introduction}
\label{sec:intro}

The term ``quantum advantage'' covers claims that differ fundamentally in what they establish. A compilation trick that removes half the ancilla qubits of a circuit, Grover's quadratic speedup with its matching query lower bound, and an exponential gap in the number of experiments needed by learners with and without quantum memory are all reported under the same name. Yet they answer different questions and fail under different restrictions. The first may be matched by a classical algorithm exploiting the same compression. The second is optimal in its query model, but its practical value may change when oracle construction is included in the cost. The third may disappear when input copies must be measured separately.

These differences motivate a more precise question: \textbf{what makes a claimed advantage possible, and what would make it disappear?} Neither the magnitude of a speedup nor the resource used by an algorithm answers this question on its own. We propose a diagnostic taxonomy that decomposes each claim into a multi-label mechanism profile, the necessity of each mechanism for the claimed separation, and the evidence supporting the classical comparison. Structural, interference, and coherent mechanisms may coexist; a mechanism's presence in a protocol does not establish that it is necessary for the advantage. The taxonomy therefore does not force claims into exclusive categories. It provides a framework for identifying what a result establishes, against which adversary, and under which resource assumptions. The mechanisms themselves are not new: quantum-memory separations, dequantization and resource theories each isolate one of them (Sec.~\ref{sec:related}). The contribution is a single procedure that applies their counterfactuals to heterogeneous claims and records the outcome in a fixed form.

The need for this distinction is especially clear in the current landscape of quantum advantage. Proved, assumption-free separations exist against restricted classical adversaries, including shallow-circuit separations~\citep{bravyi2017shallow} and sampling results~\citep{benewatts2026unconditional}. These results do not by themselves establish an advantage for a practically consequential task. Proved separations against general classical computation are known relative to oracles~\citep{aaronson2014forrelation,raz2018oracle}, including verifiable examples~\citep{yamakawa2022verifiable}. Strong conditional results can provide verification and detailed resource accounting, as in decoded quantum interferometry on optimal polynomial intersection~\citep{jordan2024dqi,khattar2025verifiable}, while remaining conditional on the strength of the best known classical attack. Meanwhile, applications in chemistry and optimization must be assessed against classical baselines that continue to improve.\footnote{We are not aware of a result that simultaneously establishes an unconditional separation, practical verifiability, and usefulness for a consequential application under a fully specified resource model.} These forms of evidence answer different questions. A taxonomy cannot settle the existence or practical value of quantum advantage, but it can make explicit which question a given result answers and which remain open.

\paragraph{Contributions.}
\begin{enumerate}[leftmargin=1.4em,itemsep=1pt]
  \item A multi-label mechanism profile---structural, interference, coherent-memory, and coherent-output---that distinguishes mechanisms used by a protocol from those necessary for its claimed separation (Sec.~\ref{sec:pillars}).
  \item A two-part entanglement-breaking test that separates loss of \emph{information} from loss of \emph{algorithmic efficiency}, together with separate registers for the adversary, evidence strength, and coherence depth (Sec.~\ref{sec:diagnostic}).
  \item A worked analysis of the hidden subgroup problem showing how mechanism profiles change when information-theoretic necessity is distinguished from algorithmic efficiency (Sec.~\ref{sec:hsp}).
  \item An audit of published claims, assessing the attribution of quantum advantage and the strength of classical comparisons (Sec.~\ref{sec:audit}).
  \item A comparison with existing classifications by evidence strength and quality of advantage, clarifying the distinct role of the proposed taxonomy (Secs.~\ref{sec:axes}--\ref{sec:related}).
\end{enumerate}


\section{Mechanism labels: a multi-label taxonomy}
\label{sec:pillars}

Table~\ref{tab:classes} summarises the mechanism labels, their characteristic counterfactuals, and the evidence needed to assess their role in a claimed advantage. The labels are not mutually exclusive. For each claim, we record (i) which mechanisms are present, (ii) which are supported as necessary for the stated separation, and (iii) the evidence, adversary, and resource model supporting that assessment. Necessity and confidence are recorded separately from mechanism presence. The witnesses listed are examples whose classical comparisons we examined, most of them in the audit (Sec.~\ref{sec:audit}).

\begin{table*}[t]
\centering
\small
\begin{tabular}{@{}p{2.2cm}p{3.1cm}p{3.1cm}p{2.9cm}p{4.1cm}@{}}
\toprule
\rr Mechanism & \rr Present when\ldots & \rr Evidence relevant to necessity & \rr Main counterfactual / attack & \rr Illustrative witnesses \\
\midrule
\rr \pillar{structural} & \rr the protocol exploits Boolean, combinatorial, or algebraic compression & \rr a matched classical comparison tests whether the same compression closes the gap & \rr dequantization; input-model conditions & \rr speedups conditional on QRAM-style or block-encoded input~\citep{kerenidis2020spectral,nghiem2025curvature} \\
\rr \pillar{interference} & \rr the protocol uses amplitude interference within a register & \rr a classical lower bound in the stated access model, i.e.\ against all dephased protocols & \rr dephasing between steps; oracle instantiation; best-known-algorithm drift & \rr own reduction~\citep{harrow2009quantum}; exact-sampling hardness~\citep{martiel2026sampling}; imported reduction~\citep{simon2026efficient}; oracle separations~\citep{aaronson2014forrelation,raz2018oracle,yamakawa2022verifiable}; imported classical bounds~\citep{grover1996fast,brassard2000amplitude,farhi2007nand} \\
\rr \pillar{coherent (memory)} & \rr the protocol jointly processes multiple input copies and is compared with one-copy-at-a-time strategies & \rr an information-theoretic lower bound against memoryless strategies supports necessity for the stated separation & \rr loss of information or loss of efficiency? & \rr \citet{aharonov2022algorithmic,huang2021information,chen2021exponential,huang2022learning,chen2026gaussian} \\
\rr \pillar{coherent (output)} & \rr the task requires a quantum output rather than classical data sufficient for the downstream task & \rr a bound against all measure-and-prepare protocols, including those using joint input measurements & \rr does the consumer need the quantum state? & \rr \citet{li2026coherent} \\
\bottomrule
\end{tabular}
\caption{Mechanism labels and evidence relevant to assessing their role. Rows are not exclusive assignments: a claim may receive multiple labels. Presence, necessity, and evidence strength are recorded separately. Within-register interference is common in quantum algorithms, so its presence alone does not establish that it is necessary for an advantage. The memory test is inapplicable to single-copy protocols; the output test is inapplicable when the task has a classical output.}
\label{tab:classes}
\end{table*}

\paragraph{Structural.}
A structural mechanism is present when a protocol exploits combinatorial, Boolean, or algebraic compression of the problem specification, for example through ancilla sharing, depth reduction by re-encoding, or compute--uncompute readout of a Born probability. Presence alone does not show that this compression explains the separation. Its counterfactual is a classical algorithm given matched access to the same structure---sampling access in place of QRAM, or the same compressed encoding. Record the mechanism as \emph{necessary} when the separation holds only relative to an input model the classical comparator does not receive; dequantization~\citep{tang2019quantum,chia2020sampling} is the characteristic demonstration. A structural label recorded as necessary is therefore a finding about the access model, not evidence of a quantum--classical separation. Where a matched-access comparison has not been made, record the mechanism as present with its role unresolved.

\paragraph{Interference.}
An interference mechanism is present when a protocol uses amplitude interference within a register. This describes most quantum algorithms and is not, by itself, evidence of advantage. Its counterfactual is dephasing: decohere the register in the computational basis after every oracle call or elementary step. A fully dephased protocol is a classical randomized algorithm with the same access, so a proved classical lower bound in that access model is exactly the statement that the separation does not survive the loss of within-register coherence. The test ranges over all dephased protocols, not over the dephased version of the algorithm under study; dephasing one algorithm and watching it fail establishes nothing (Sec.~\ref{sec:hsp}). Record interference as \emph{necessary} only with such a bound, and only in the access model where it is proved. A bound that holds unless a named complexity-class conjecture fails, such as $\mathrm{BQP}\not\subseteq\mathrm{BPP}$ or the non-collapse of the polynomial hierarchy, supports \emph{necessary, conditional}. A bound that rests instead on the classical hardness of a specific problem---factoring, the discrete logarithm, decisional Diffie--Hellman, learning with errors---is best-known-algorithm evidence however long the problem has been studied, and supports \emph{candidate}. A reduction to such a problem is recorded in the evidence register; it does not change the status. Where only selected classical methods have been beaten, record interference as a \emph{candidate} and the comparison as benchmark evidence. Necessity in a query model does not transfer to an instantiated oracle: a query advantage is an end-to-end advantage only when the cost of constructing and accessing the oracle is included or explicitly excluded from the claim.

\paragraph{Interference under the stabilizer restriction.}
Because a dephased protocol is a classical randomized algorithm, \emph{interference necessary} is by construction equivalent to a classical lower bound in the stated access model. On its own the label therefore adds little to the evidence register; what it lacks is the non-classical resource behind the bound. A second counterfactual supplies it: restrict the protocol to Clifford gates, Pauli measurements, and Clifford gates conditioned on classical bits, including earlier measurement outcomes. Such circuits interfere amplitudes yet can be simulated in polynomial time on a probabilistic classical computer (the Gottesman--Knill theorem~\citep{gottesman1998heisenberg}), so interference is not sufficient for a separation from polynomial-time classical computation on a classical input. The restriction splits the label. If the separation survives it, the separation can hold only against a restricted adversary or in a query model, and magic is not its source. For the shallow-circuit separations the witness is nonlocality; for Clifford-only query separations, such as Bernstein--Vazirani with a linear phase oracle, no such witness has been identified. For each input, the hidden-linear-function circuit applies Clifford gates to its data register, so the problem is solvable classically through the Gottesman--Knill theorem, and the separation from constant-depth classical circuits rests on graph-state nonlocality~\citep{bravyi2017shallow}; the one-dimensional version is a depth-four circuit of classically controlled Clifford gates, and its noise-tolerant three-dimensional encoding remains a classically controlled Clifford circuit~\citep{bravyi2020noisy}. If the separation does not survive, magic is necessary, as it is for any separation from polynomial-time classical computation on a classical input without an oracle. Record the outcome as \emph{stabilizer-robust} or \emph{magic-necessary}. Magic and contextuality are related but not interchangeable. Contextuality with respect to stabilizer measurements is necessary for universal quantum computation by magic-state distillation, and for qudits of odd prime dimension it coincides with the known distillation boundary; for qubits, its mere presence is not sufficient, since even the maximally mixed state violates a noncontextuality inequality built from stabilizer measurements~\citep{howard2014contextuality}. The register therefore names the counterfactual applied rather than a resource.

\paragraph{Coherent.}
A coherent mechanism is implicated when restricting quantum information to a classical record changes what can be achieved. Two distinct restrictions are relevant here. The literature uses terms such as ``incoherent'', ``measure-first'', and ``entanglement-breaking'' for settings that may impose different restrictions. We therefore record memory and output separately, without assuming that either is the sole or primary mechanism.

\emph{Coherent (memory).}
This mechanism is present when the protocol jointly processes multiple copies of the input, $\rho^{\otimes m}$ with $m\ge2$, and the comparison restricts the adversary to measuring each copy separately and post-processing the resulting classical records. The canonical examples include sample-complexity separations between learners with and without quantum memory~\citep{chen2021exponential,huang2022learning}. The output is classical. To assess necessity, determine whether the restriction causes an information-theoretic loss or only an efficiency loss, as discussed in Sec.~\ref{sec:diagnostic}.

\emph{Coherent (output).}
This mechanism is present when the task requires a quantum output, such as a state or an operation applied to a downstream register, and the comparison is against measure-and-prepare protocols. Such an adversary may measure all input copies jointly, but must prepare its output from a classical record. The relevant restriction is therefore on the output, not necessarily on joint processing of the inputs. \citet{li2026coherent} prove separations for quantum purity amplification with principal-eigenstate targets, with $O(1/\varepsilon)$ copies compared with $\Omega(d/\varepsilon)$, and for density-matrix exponentiation. Joint processing of input copies is not inherently required by this output restriction, so the coherence depth (Sec.~\ref{sec:diagnostic}) may be $1$. The relevant counterfactual concerns the consumer: if downstream use requires only expectation values or other classical data, rather than the state itself, the quantum-output separation may not apply.

These coherent subtypes are distinct tests, not exclusive destinations. The memory test is inapplicable to single-copy protocols, while the output test is inapplicable to tasks with classical outputs. A claim may receive both labels if both restrictions are relevant, but each necessity assessment must identify the restriction addressed by its supporting lower bound. A protocol that produces a classical output is not, for that reason alone, evidence of a coherent-output separation.


A claimed advantage may involve multiple mechanisms or several multiplicative contributions to a speedup. Record each mechanism's presence and necessity separately rather than assigning the claim a single label. For example, \citet{incudini2026practical} report ``about a cubic polynomial asymptotic advantage over previous quantum-walks, resulting in a total sixth-degree polynomial queries speedup compared to the best classical walk''. The quadratic contribution comes from the Szegedy walk, which converts a spectral gap $\delta$ into a phase gap of order $\sqrt{\delta}$ and uses within-register interference. The authors attribute the cubic contribution to a Hamiltonian-simulation proposal that enlarges the Markov-chain gap itself, an effect they note can also be obtained by applying the proposal sequentially, one measured step at a time. That factor is an exponent fitted to spectral gaps of Sherrington--Kirkpatrick instances with $n=3$--$10$ spins, not a bound. This indicates that the cubic contribution does not require coherent processing across those steps. It does not, by itself, establish the contribution's status against all classical alternatives. The profile should therefore record within-register interference as present, distinguish the two speedup contributions, and assess their necessity and classical comparison separately. Bounded memory or bounded coherence depth should likewise be recorded as resource restrictions, not automatically relabelled as coherent-memory or coherent-output mechanisms unless the corresponding test is met.


\section{The diagnostic and a two-part entanglement-breaking test}
\label{sec:diagnostic}

The entanglement-breaking test asks whether a claimed separation survives when specified forms of quantum coherence are removed. Its interpretation depends on which coherence is removed and which adversary is allowed. Applied \emph{within} a copy, by measuring the input before within-register interference, it removes a resource used by a broad range of quantum algorithms, including Fourier-sampling algorithms. Failure under this test therefore does not, by itself, establish that interference is necessary for the separation. Only a lower bound over all dephased protocols, which is a classical lower bound, does. Applied \emph{across} copies, by restricting measurements to act on one copy at a time, it is vacuous for single-copy protocols. We therefore use a two-part test to construct a multi-label mechanism profile, not to force a single verdict.

\begin{enumerate}[leftmargin=1.4em,itemsep=1pt]
  \item \textbf{Describe the protocol and task.} Record whether the protocol processes multiple input copies jointly, whether its required output is quantum or classical, and the input-access model. Also specify the resource used to state the separation, such as copies, queries, time, or precision. These facts determine which tests apply.

  \item \textbf{Test coherent-memory relevance.} Compare the protocol with strategies that measure each input copy using an arbitrary single-copy POVM and then post-process the classical outcomes. Record \emph{information-theoretic necessity} only when a proved lower bound shows that this restriction increases the required sample complexity for the stated task. If the samples still suffice but no efficient classical post-processing method is known, record cross-copy coherence as \emph{used by the known algorithm; necessity unresolved}. This indicates an algorithmic gap, not an established information-theoretic gap.

  \item \textbf{Test coherent-output relevance.} If the task requires a quantum output, compare the protocol with measure-and-prepare channels, allowing the adversary to perform a joint measurement on all input copies before preparing the output from a classical record. Record coherent-output necessity only when a proved bound establishes a separation in a specified resource, such as copies or precision. This test does not establish a coherent-output advantage when the required output is classical.

  \item \textbf{Record interference.} Identify whether the protocol uses within-register amplitude interference. Separately record whether a proved classical lower bound in the stated access model establishes the claimed separation. Because a dephased protocol is a classical randomized one, such a bound establishes that within-register coherence is necessary in that access model; it does not show that no other mechanism also contributes. A bound that holds unless a named complexity-class conjecture fails supports \emph{necessary, conditional}; one resting on the hardness of a specific problem, or that only beats selected classical methods, supports \emph{candidate}, with the comparison recorded as best-known or benchmark evidence. Then apply the stabilizer restriction (Sec.~\ref{sec:pillars}) and record the separation as stabilizer-robust or magic-necessary.

  \item \textbf{Record structural compression.} Identify any Boolean, combinatorial, or algebraic compression used by the protocol. Mark the mechanism as necessary for the separation only when a counterfactual comparison supports that conclusion. Otherwise, record it as present, with its causal role unresolved.

  \item \textbf{Allow multiple labels.} Retain every mechanism supported by the evidence. For each mechanism, record its status as \emph{present}, \emph{necessary}, \emph{necessary, conditional}, \emph{candidate}, or \emph{unresolved}, together with a short justification and the adversary or comparison to which the claim applies. The presence of one mechanism does not exclude the contribution of another.
\end{enumerate}


Where the relevant quantities are defined, record the coherence depth as the pair
$(m_{\mathrm{info}},m_{\mathrm{alg}})$, where
$m_{\mathrm{info}}$ is the minimum number of copies that must be processed jointly to attain the stated task's performance, and
$m_{\mathrm{alg}}$ is the minimum number jointly processed by the algorithm under study. These minima must be defined relative to the task, target performance, access model, and allowed strategies. Within-register interference can occur at depth $(1,1)$. A confirmed memory separation can require $m_{\mathrm{info}}\geq 2$ and $m_{\mathrm{alg}}\geq 2$. If $m_{\mathrm{alg}}>m_{\mathrm{info}}$, the known algorithm uses more joint processing than the task is known to require; this alone does not prove that a more efficient algorithm exists. Coherence depth is a diagnostic for the memory test, not a general measure of quantum advantage: a coherent-output witness can have depth $1$.

Independently record the comparator, the status of the evidence, the resource measured, and the practical quality of the result. Specify whether the classical comparator covers all algorithms in a defined model, a restricted algorithm class, or selected implementations. Classify the evidence as proved, imported from another result, conditional on a named complexity-class conjecture, relative to the best-known algorithm, or empirical, and state whether the resource is queries, copies, time, space, precision, or end-to-end cost. The two middle classes differ: BQP-hardness rests on a conjecture the community treats as a standing assumption, while best-known-algorithm comparisons can be overturned by a single new algorithm. Hardness assumptions on specific problems, such as factoring, the discrete logarithm or learning with errors, belong to the second class however long they have stood, including when a separation is proved by reduction to them. Record practical quality separately, including relevant implementation costs and application constraints. A comparison with the best known algorithm is benchmark evidence, not a lower bound (Sec.~\ref{sec:hygiene}).


\section{A worked case: the hidden subgroup problem}
\label{sec:hsp}

The hidden subgroup problem (HSP) illustrates why mechanism profiles must distinguish interference from coherent memory. The same problem family can produce different profiles depending on the group and the available access model. Given oracle access to a function on a group $G$ that is constant exactly on the cosets of an unknown subgroup $H$, the standard quantum approach prepares coset states and processes them using the quantum Fourier transform.

Measure a coset state in the computational basis before the Fourier transform and the outcome is a uniformly random group element, independent of the hidden subgroup. This destroys the information exploited by standard Fourier-sampling algorithms, including those for Simon's and Shor's problems. The test therefore demonstrates the importance of preserving quantum processing within a register, but it does not establish that interference is necessary for a particular separation. Nor does it distinguish coherent-memory from coherent-output mechanisms.

Table~\ref{tab:hsp} records the coherence-depth pair $(m_{\mathrm{info}},m_{\mathrm{alg}})$ for four groups. Here $m_{\mathrm{info}}$ refers to how many copies must be measured jointly for the hidden subgroup to be determinable from polynomially many copies overall, and $m_{\mathrm{alg}}$ to the number of copies jointly processed by the cited efficient algorithm, where known. The profile distinguishes information-theoretic memory requirements from cases where joint processing is used by the known algorithm but is not known to be necessary for extracting enough information.

\begin{table*}[tbp]
\centering
\caption{Joint measurement in the hidden subgroup problem, recorded as the coherence-depth pair of Sec.~\ref{sec:diagnostic}. $m_{\mathrm{info}}$ is the number of copies that must be measured jointly to determine $H$ from polynomially many copies overall; $m_{\mathrm{info}}=1$ means single-register measurements suffice information-theoretically. $m_{\mathrm{alg}}$ is the number of copies jointly processed by the best efficient algorithm cited here. Sources: \citet{childs2025lecture}, Chapters 6 and 12--14, and the papers cited in the text.}
\label{tab:hsp}
\small
\begin{tabular}{@{}p{0.35\columnwidth}p{0.20\columnwidth}p{0.30\columnwidth}p{0.95\columnwidth}@{}}
\toprule
Group & $m_{\mathrm{info}}$ & $m_{\mathrm{alg}}$ & Mechanism profile \\
\midrule
abelian; normal $H$ & 1 & 1 & interference present; no memory requirement \\
\addlinespace
Heisenberg $\mathbb Z_p^2\rtimes\mathbb Z_p$ & 1 & 2 & interference present; memory-efficiency provisional \\
\addlinespace
dihedral $D_N$ & 1 & $2^{O(\sqrt{\log N})}$ & interference present; memory-efficiency provisional \\
\addlinespace
symmetric $S_n$ & $\Omega(n\log n)$ & no efficient algorithm known & interference present; memory-information$^\ast$ \\
\bottomrule
\end{tabular}

\par\vspace{2pt}
{\footnotesize\raggedright $^\ast$Information-theoretically, joint measurements are required in the stated setting; no efficient algorithm is known.\par}
\end{table*}

For abelian groups, and for normal hidden subgroups of any group, Fourier sampling on one register at a time determines $H$ with $O(\log |G|)$ repetitions~\citep[Chapters 6 and 12]{childs2025lecture}. The profile therefore records interference as present, with no cross-copy memory requirement. Any claim about the source of the overall separation must still be supported by the relevant classical comparison. For Simon's problem that comparison exists: a classical randomized algorithm needs exponentially many queries~\citep{simon1997power}, so interference is necessary in the query model in the sense of Sec.~\ref{sec:pillars}.

For the Heisenberg group, one sample does not determine $H$, but random measurements on individual registers of $\mathrm{poly}(\log p)$ samples do, so $m_{\mathrm{info}}=1$. The difficulty is extracting the answer efficiently from these outcomes. A joint measurement on two copies provides an efficient solution~\citep[Chapter 14]{childs2025lecture}. The profile consequently records joint processing as used by the known algorithm, while treating its necessity for efficient computation as an algorithmic question rather than an information-theoretic one.

For the dihedral group, single-register measurements produce distributions that are statistically distinguishable for different hidden reflections, so polynomially many samples suffice in principle. However, efficiently extracting the hidden reflection from these samples remains difficult. Kuperberg's sieve~\citep{kuperberg2005subexponential} solves the problem in time $2^{O(\sqrt{\log N})}$ by repeatedly combining registers. Its joint processing is central to the known algorithm, but that fact alone does not establish an information-theoretic need for joint measurements. The optimal-measurement threshold at $\Theta(\log N)$ copies~\citep{bacon2005optimal} concerns the total number of copies required, not the number that must be measured jointly in a single operation. These are distinct resource questions.

For the symmetric group, single-register strong Fourier sampling fails, even with arbitrary single-copy POVMs~\citep{moore2005symmetric}. \citet{hallgren2006limitations} prove that no measurement on $o(n\log n)$ coset states extracts a polynomial amount of information about the hidden subgroup, so any polynomial-time algorithm operating on coset states must measure $\Omega(n\log n)=\Omega(\log|G|)$ of them jointly, matching the information-theoretic upper bound. The hard instances are the order-two subgroups arising from the standard reduction of graph isomorphism. Among the four examples considered here, this is the case that supports an information-theoretic coherent-memory label. The claim is relative to the stated measurement model and task, rather than a universal statement about every formulation of the symmetric-group HSP.

The within-copy test does not distinguish these examples: all use quantum Fourier processing, but its presence does not establish that interference is the necessary cause of a separation. The cross-copy test is more discriminating. For abelian groups and normal subgroups, one-copy-at-a-time processing suffices. For the Heisenberg and dihedral cases, the known efficient algorithms use joint processing, while single-copy measurements can provide sufficient information; the evidence therefore supports a provisional memory-efficiency profile, not a memory-information separation. For the symmetric-group case, the cited lower bounds support an information-theoretic memory requirement. Kuperberg's sieve illustrates the distinction: joint processing is important to the known algorithm's efficiency, but this is not by itself evidence that joint processing is needed to obtain enough information.

The Abelian \emph{state} HSP provides a useful contrast. Here, the input is a quantum state with a hidden symmetry rather than oracle access to a group function. \citet{liu2026power} establish matching worst-case bounds for a constructed family of padded Simon-type states. Given copies of the state, $\Theta(\log(|G/H|)/\varepsilon)$ copies are necessary and sufficient, even when arbitrary collective measurements are allowed. Collective processing therefore provides no asymptotic copy-complexity improvement in this setting. With access to the preparation unitary and its inverse, $\Theta(\log(|G/H|)/\sqrt{\varepsilon})$ queries are necessary and sufficient. The quadratic improvement is attributed to coherent access to the preparation circuit, through amplitude amplification of informative Fourier samples.

In the present taxonomy, this is an interference mechanism rather than a coherent-memory mechanism: the advantage comes from reversible access to the preparation process, not from jointly measuring several independently supplied copies. It also illustrates a different kind of separation from the standard quantum-versus-classical comparison. The two bounds compare different quantum access models, and the stronger model includes a preparation unitary and its inverse. The result therefore establishes an access-model advantage, not a separation from classical algorithms.

Before attributing a gain to joint processing of copies, ask whether reversible access to a single preparation could achieve the same improvement. Then ask whether a lower bound against collective measurements establishes that the copy-based task genuinely requires joint processing. These questions distinguish gains from coherent access to a preparation process, gains in algorithmic efficiency, and gains that require joint measurements for information-theoretic reasons.


\section{Auditing published claims}
\label{sec:audit}

We applied the procedure of Sec.~\ref{sec:diagnostic} to fifty papers. All had been labelled in an earlier screening that read summaries of each paper against a single entanglement-breaking test. For each paper, we then read the paper itself and asked four questions: what exactly is claimed, and against which comparator; whether a lower bound is proved in the paper, imported from another, or absent; whether the protocol is single-copy, multi-copy, or has a quantum output; and which mechanism labels and necessity statuses are supported. The labels are ours. Most of these authors make no claim in our vocabulary, and a changed label is a finding about the diagnostic, not a criticism of the paper.

\begin{table}[t]
\centering
\small
\setlength{\tabcolsep}{4pt}
\begin{tabular}{@{}lcccc@{}}
\toprule
First label & $N$ & Changed & Reason fixed & Stands \\
\midrule
coherent, confirmed & 14 & 9 & 3 & 2 \\
interference & 16 & 7 & 5 & 4 \\
coherent, benchmark & 6 & 6 & --- & --- \\
coherent, other & 3 & 2 & --- & 1 \\
\midrule
total & 39 & 24 & 8 & 7 \\
\bottomrule
\end{tabular}
\caption{Re-reading the provisional labels retained by the first screening. These counts audit the initial screening labels and do not represent mutually exclusive final classes. ``Reason fixed'' means that the initial label remains defensible but its justification needed correction. Examples include an imported bound, a bound against quantum rather than classical algorithms, or a claim reassigned from the memory test to the output test. Eleven further initial labels, already revised by the screening, were re-checked separately against the papers, and three of those revisions were themselves corrected. The revised procedure permits multiple mechanism labels per claim.}
\label{tab:audit}
\end{table}

Table~\ref{tab:audit} reports the audit of the first screening's provisional labels. Of the 39 labels it had retained, 24 were revised, while eight additional labels remained defensible only with a corrected justification. These counts describe changes to the initial assignments, not mutually exclusive final categories under the multi-label procedure. The revisions reveal several recurring problems: mistaking a protocol ingredient for a demonstrated separation, confusing quantum optimality with a classical lower bound, applying a theorem against the wrong adversary, treating benchmark comparisons as lower bounds, and overlooking restrictions in the input-access model.

A first problem arises when a mechanism label is triggered by a coherence-related feature of a protocol rather than by a separation that requires it. One example is a two-copy estimator in a Bayesian framework whose only proved gap is against a naive single-copy estimator~\citep{jager2026provable}. Another is the classical-shadow protocol, which serves as the memoryless baseline rather than its opposite~\citep{huang2020predicting}. A further six papers on entangling dynamics and photonic kernels were labelled coherent based on the colloquial use of ``coherent''. These protocols operate on one register and output classical data, so both coherent-memory and coherent-output tests are vacuous. Three of them make no quantum-versus-classical claim on reading. For the other three, the evidence shows that selected classical methods fail, and two of the papers say explicitly that this is not a bound on all classical methods. For example, the scaling analysis of~\citet{king2024beyond} ``is not an intrinsic lower bound for all classical methods''. In these cases, within-register interference is present, but the classical comparison is benchmark-level rather than a proved lower bound~\citep{king2024beyond,sun2026crossover,leviatan2026resolving}. The profile should therefore record interference as a candidate mechanism, or record no advantage claim where the paper makes none.

A related classification issue occurs in a noisy-data learning separation~\citep{danaci2026evidence} that is single-copy on both sides. Its asymptotic bound is imported from a separation between task-adapted and \emph{measure-first} learners~\citep{gyurik2023limitations}. In that result, ``measure-first'' does not mean measuring one copy at a time: the learner may measure all copies jointly and is restricted to a polynomial-time measurement strategy chosen independently of the target concept. This supports a separation from a fixed classical sketch, not a coherent-memory separation.

A second problem is the substitution of a quantum optimality bound for the classical bound needed to establish a quantum-classical separation. Grover's algorithm~\citep{grover1996fast} is optimal among quantum algorithms by an $\Omega(\sqrt N)$ bound~\citep{bennett1997strengths}, but this bound does not establish the classical search lower bound of $\Omega(N)$ that supports interference as necessary in the query model. The same issue arises for the NAND tree, where the cited lower bound is quantum while the classical $N^{0.753}$ result is separately relevant~\citep{farhi2007nand}, and for a matrix-ODE solver, where the solver's own bound is quantum while classical hardness is inherited from BQP-completeness~\citep{simon2026efficient}. In these cases, the mechanism label can remain defensible, but the original justification requires correction. Where only quantum optimality is established and no classical hardness result is invoked, the claim that interference is necessary for a quantum-classical separation is unsupported, even if the protocol uses interference.

A third problem is applying a valid theorem to a different adversary from the one required by the mechanism test. \citet{li2026coherent} bound measure-and-prepare protocols that may measure all input copies jointly, which is why coherent-memory and coherent-output require distinct tests. \citet{zhao2026exponential} prove an unconditional separation against classical machines with bounded memory on classical data streams: polylogarithmic quantum memory against any classical memory of size $O(N^{c})$, $c<1$, at equal sample count. The resource axis is space, and no copies of a quantum state are involved. \citet{benewatts2026unconditional}, extending~\citet{bravyi2017shallow}, bound classical circuits of bounded fan-in and limited randomness, and explicitly state that the bound ``applies only to NC0 circuits with at most $n + n^\delta$ random input bits''. \citet{nguyen2026symmetries} establish a separation between quantum and classical statistical queries, where the classical comparator is restricted. Finally, \citet{schuster2024random} prove indistinguishability against quantum adversaries under a conjecture on learning with errors and import their quantum-classical separation from memory results. Each is a genuine result, but none by itself establishes the broader mechanism label assigned in the initial screening. The comparator and resource restriction must match the claim being made.

A fourth issue is the absence of a lower bound. A polynomial gain, at most quadratic, over one classical algorithm, without a lower bound, supports a candidate advantage rather than a proved separation. Examples include the quantum-walk search framework compared with the analogous classical walk~\citep{magniez2007search}, the quantized walk on which it builds~\citep{szegedy2004spectra}, and a quantum Bayesian-network planner compared with classical rejection sampling~\citep{cunha2026quantum}. Similarly, a multi-copy protocol without a single-copy lower bound~\citep{yuan2026qpca} uses cross-copy coherence in the known algorithm, but its necessity status remains unresolved. Density-matrix exponentiation~\citep{lloyd2013quantum} does not itself prove a separation; a later result proves a separation for density-matrix exponentiation itself against all measure-and-prepare protocols, $O(T^2/\varepsilon)$ copies against $\Omega(\sin^2(T/2)\,d/\varepsilon)$~\citep{li2026coherent}.

Benchmark comparisons also change as classical algorithms improve. QAOA was reported to outperform the best known classical algorithm with a provable guarantee for Max-Cut on 3-regular graphs of girth at least 16, at depth $p\ge7$~\citep{farhi2025lower}. About eighteen months after that report first appeared, a classical local algorithm exceeded the QAOA guarantees at every depth $p\ge4$ for which they are available~\citep{morone2026whither}. A related claim for higher degree and Max-$k$-Cut~\citep{apte2026maxkcut} has not been tested by that algorithm. Such developments do not invalidate a proved lower bound, but they can change the standing of a best-known-algorithm comparison. The comparator and the date or version of the benchmark therefore matter to the interpretation of empirical or algorithm-relative claims.

Input access is another source of apparent advantage. Two speedups rely on access models such as QRAM-style sampling or block-encoded distances that are not provided to the classical comparator~\citep{kerenidis2020spectral,nghiem2025curvature}. These cases motivate the structural label, but the label does not itself establish a quantum-classical separation. The input model and the cost of preparing or accessing the data must be included in the comparison, and dequantization provides a natural counterfactual to test whether the same structure can be exploited classically.

Under the revised diagnostic, three memory witnesses have their own proofs~\citep{chen2021exponential,huang2022learning,chen2026gaussian}, although the last applies only to cold states. We also include the earliest such result, outside the audited sample: \citet{aharonov2022algorithmic} prove exponential query lower bounds against \emph{adaptive} experimentalists who measure each oracle output before the next call. The audit also identifies one output witness~\citep{li2026coherent}, two interference witnesses with their own conditional classical hardness results, namely a BQP-completeness reduction~\citep{harrow2009quantum} and exact-sampling hardness, with approximate hardness conditional on two conjectures~\citep{martiel2026sampling}, and additional interference witnesses whose classical bounds are imported. Two further memory witnesses import or restate their bounds and hold in the worst case only. These examples differ in evidence strength and comparator, so they should not be treated as interchangeable instances of a single kind of advantage.

The audit also shows how easily a summary or secondary description can overstate what a paper establishes. Some initial labels rested on paraphrases that attributed to a paper claims it does not make, such as a separation that ``vanishes if entanglement between input copies is broken'' or an exponential sample cost for single-copy estimation. The first question in the procedure, \emph{which theorem, against which adversary}, is a simple safeguard. Answering it requires reading the theorem statement and its assumptions, rather than relying on a summary of the paper.


\section{Mechanism, evidence, and quality}
\label{sec:axes}

A quantum-advantage claim should be reported along three conceptually distinct coordinates. The first is its multi-label mechanism profile, including whether each mechanism is present, necessary, or unresolved. The second is the strength of the evidence supporting the classical comparison. \citet{lin2026quantum} organize quantum applications into four levels: highly structured problems with provably expensive classical cost (I); unitary dynamics, sampling, and learning, where classical cost is empirically expensive (II); quantum properties and effective dynamics, such as ground-state energies, where input and output costs remain unsettled (III); and problems with strong classical baselines or limitations arising from input and output costs (IV). The third coordinate is the quality of the claimed advantage. \citet{huang2026vast} assess this through five keystones. These coordinates address different questions and should be recorded separately, although they need not be statistically or mathematically independent. Table~\ref{tab:axes} illustrates how this approach distinguishes claims that a single mechanism label would otherwise conflate.

\begin{table}[t]
\centering
\small
\begin{tabular}{@{}p{2.4cm}p{2.8cm}p{1.9cm}@{}}
\toprule
\rr Result & \rr Mechanism profile & \rr Classical side \\
\midrule
\rr factoring~\citep{shor1995polynomial} & \rr interference candidate (explicit input); necessary for period finding (query) & \rr best known \\
\rr linear systems~\citep{harrow2009quantum} & \rr interference necessary, conditional & \rr BQP $\neq$ BPP (own reduction) \\
\rr exact RCS-type sampling~\citep{martiel2026sampling} & \rr interference necessary, conditional & \rr PH non-collapse \\
\rr unstructured search~\citep{grover1996fast} & \rr interference necessary (query) & \rr query bound, imported \\
\rr quench dynamics~\citep{king2024beyond} & \rr interference candidate & \rr benchmark \\
\rr learning from experiments~\citep{huang2022learning} & \rr coherent-memory necessary & \rr information-theoretic \\
\rr coherent inference~\citep{li2026coherent} & \rr coherent-output necessary & \rr information-theoretic \\
\rr verifiable search~\citep{yamakawa2022verifiable} & \rr interference necessary (random oracle) & \rr query bound \\
\rr shallow circuits~\citep{bravyi2017shallow} & \rr interference necessary against NC$^0$ & \rr circuit-class bound \\
\rr dequantized recommendation~\citep{tang2019quantum} & \rr structural necessary & \rr matched (dequantized) \\
\bottomrule
\end{tabular}
\caption{Examples illustrating the distinction between a mechanism profile and the evidence supporting the classical comparison. These entries describe different kinds of evidence and comparator, not a common numerical scale of evidential strength. \citet{lin2026quantum} place factoring at Level~I and Hamiltonian simulation, random circuit sampling, and learning with quantum memory at Level~II; the last placement is discussed in the text. Under the stabilizer restriction (Sec.~\ref{sec:pillars}), the shallow-circuit entry is stabilizer-robust, while the linear-systems and exact-sampling entries are magic-necessary.}
\label{tab:axes}
\end{table}

This separation has two important consequences. First, a mechanism label is not an endorsement of the strength or practical significance of a claim. A claim can have a clear mechanism profile while its classical comparison remains weak. Lin and Wiebe place ground-state and materials problems that dominate many application roadmaps at Level~III, citing strong approximate classical methods and unsettled input costs. That classification is distinct from the mechanisms used by the corresponding quantum algorithms.

Second, the evidence taxonomy permits a weaker classical comparison than the criterion used here to establish interference as necessary for a separation. Lin and Wiebe describe a classical cost as ``provably expensive'' when the problem is hard under complexity-theoretic conjectures or relative to the best-known classical algorithms. The latter criterion is a benchmark comparison, not a lower bound against the relevant class of classical algorithms. Factoring is placed at Level~I partly on this basis. In the present taxonomy, interference is necessary for period finding in the query model, but for factoring an explicit integer it remains a candidate: its classical hardness is a problem-specific assumption supported by the best-known algorithm, not a complexity-class conjecture. The evidence register makes that qualification explicit. A Level~I classification therefore does not, by itself, establish that interference is necessary for the separation.

Third, the two schemes can rank the same result in opposite directions. Lin and Wiebe place learning with quantum memory at Level~II, whose classical cost is ``empirically expensive'', whereas Table~\ref{tab:axes} records the learning-from-experiments separation~\citep{huang2022learning} as information-theoretic. Both placements are defensible because they answer different questions. Their levels are assigned to categories, each ``determined by the most compelling application'' in it, so learning shares its level with Hamiltonian simulation and random circuit sampling. More importantly, the lower bound behind the separation~\citep{chen2021exponential,huang2022learning} is unconditional but holds against a restricted adversary: a learner that measures each copy, or each use of the process, separately and keeps only classical records. It says nothing about a classical computer given a classical description of the system, which is the comparator in a hierarchy built for problems with classical inputs and outputs. Lin and Wiebe state this scope directly: such advantages ``primarily stem from the quantum nature of the input data and the availability of quantum memory'', and the exponential separation assumes ``zero knowledge about the quantum system being learned, which is often not the case in physical applications''. An information-theoretic bound is the strongest support this taxonomy records for a coherent-memory label, yet it establishes necessity only within the stated access model. Prior knowledge of the system, or a classical description of it, places the comparison outside that model. Reporting the access model alongside the label prevents an unconditional bound from being read as evidence of a practical advantage.

The five keystones proposed by \citet{huang2026vast}, namely predictable, typical, robust, verifiable, and useful, assess a further dimension of the claimed advantage. The memory separations discussed above are typical and robust in theory, but concern sample complexity rather than running time. A quench-dynamics advantage may have practical value while remaining difficult to verify by construction. These assessments complement, rather than replace, the mechanism profile and the evidence register. Taken together, the three coordinates describe which mechanisms are involved, how strongly the classical comparison is supported, and what practical value the claimed advantage may offer.


\section{Relation to existing classifications}
\label{sec:related}

The value of a taxonomy depends on what it distinguishes from existing classifications. Six bodies of work are particularly relevant: quantum-memory separations, dequantization, trainability and simulability, resource theories, frameworks for evaluating advantage, and algorithm catalogues. The aim here is not to replace these approaches, but to clarify which questions the mechanism profile answers and how it can be used alongside them.

The closest prior work is the line of results showing that learners able to store copies of a quantum state and measure them jointly can require exponentially fewer samples than learners that measure each copy and post-process classically. These separations have been established for shadow tomography, purity testing, and the detection of symmetries in dynamics~\citep{chen2021exponential}, with demonstrations on hardware for tens of qubits~\citep{huang2022learning}; see also~\citet{aharonov2022algorithmic,huang2021information}. \citet{huang2021information} also prove a converse bound for this setting: for \emph{average} prediction error, a learner that measures each output separately needs at most $O(mN_Q/\varepsilon)$ uses of the process, where $N_Q$ is the quantum learner's count and $m$ the output size. Thus, an average-error claim cannot establish an exponential memory separation on its own. Constant local noise without error correction damps the worst-case separation exponentially, although the base can remain close to one at small noise rates~\citep{chen2022nisq}.

Our coherent-memory label describes the mechanism exhibited by these results, and the across-copy test formalizes the restriction to measurements on individual copies followed by classical post-processing. We do not claim a new boundary for established memory separations. The results of~\citet{li2026coherent}, by contrast, concern a different adversary: measure-and-prepare protocols are allowed to perform joint measurements on all input copies, but must prepare the output from a classical record. These results support the coherent-output label. Distinguishing this adversary from one restricted to measuring copies individually is essential to avoiding the misplaced coherent labels identified in our audit. The additional distinction we introduce is between what a restriction costs in information and what it costs in the efficiency of a known algorithm. This distinction is not needed to characterize memory separations already established by information-theoretic lower bounds, but becomes important when the diagnostic is applied to claims without such bounds. The hidden subgroup problem (Sec.~\ref{sec:hsp}) illustrates how information and efficiency can come apart, while the abelian state version provides the opposite case, where joint processing of copies provably offers no asymptotic improvement~\citep{liu2026power}.

Dequantization provides a second point of comparison. Quantum-inspired classical algorithms~\citep{tang2019quantum,chia2020sampling} distinguish speedups that survive classical sampling access to the same data from those that depend on a more powerful input model. We do not attempt to reproduce that classification. Instead, the structural label identifies cases in which dequantization is a natural counterfactual. An apparent coherent-mechanism advantage can arise from an input-access advantage, such as QRAM-style sampling access, rather than from the computation itself. The mechanism profile therefore complements dequantization by recording structural compression and by distinguishing the use of a resource from evidence that the resource is necessary for the claimed separation.

Results on barren plateaus, classical simulability, and concentration impose further constraints on the practical relevance of coherent mechanisms. Provable absence of barren plateaus tends to accompany classical simulability~\citep{cerezo2025barren}, while quantum kernels can concentrate exponentially under the conditions identified by~\citet{thanasilp2024exponential}: high expressivity of the data embedding, global measurements, entanglement, and noise. Together, these results identify two potential limitations: insufficient expressivity can leave a model classically simulable, while excessive expressivity can make learning difficult. The geometric difference introduced by~\citet{huang2021power} also provides a data-dependent prescreen. We treat these findings as constraints on where coherent mechanisms may be useful in practice, not as part of the definition of those mechanisms.

Resource theories offer another distinct perspective. They quantify which resources, such as entanglement, coherence, or magic, support a task, typically through monotones~\citep{chitambar2019resource}. Our mechanism labels are neither monotones nor rankings of quantum states. They characterize a claim through the counterfactuals under which its proposed advantage would fail. Two protocols that consume the same amount of entanglement can therefore receive different profiles: removing a resource may destroy the information available to one protocol while affecting only the efficiency of another.

The quality framework of~\citet{huang2026vast} addresses a further question. Its five keystones, predictable, typical, robust, verifiable, and useful, assess the quality of a claimed advantage, while its four realms (computation; learning and sensing; cryptography, communication, and strategic games; and space) group advantages by domain, and these differ in what they rest on, from unproven complexity assumptions to the laws of physics. These dimensions complement the mechanism profile, which asks where the claimed advantage comes from. Learning-from-experiment separations can receive a coherent-memory label, dequantized data-loading speedups a structural label, and single-copy results such as Shor's algorithm and random circuit sampling an interference label when the relevant classical comparison supports it. An interference-based advantage may be predictable yet atypical, while a coherent advantage may be robust in theory but difficult to verify in practice. Within our scheme, decoded quantum interferometry~\citep{jordan2024dqi,khattar2025verifiable} is recorded as an interference candidate because its classical comparison is with the best-known attack rather than a proved bound.

The relationship between mechanism and application is also relevant to the critique by~\citet{schuld2022advantage}, who question whether advantage should organize quantum machine learning at all and observe that problems studied in quantum computing are typically ``very structured'', often ``exhibiting a periodic structure that can be exploited by interference''. The mechanism profile makes this observation more precise by identifying interference as a candidate or necessary mechanism according to the available evidence. The provable separations for classical data~\citep{liu2020rigorous,pirnay2022density} are interference advantages inherited from discrete-logarithm structure. By the rule applied to factoring, interference is a \emph{candidate} in both: their classical hardness is proved by reduction to the discrete logarithm~\citep{liu2020rigorous} and to the decisional Diffie--Hellman problem~\citep{pirnay2022density}, which are best-known-algorithm assumptions. The reductions are recorded in the evidence register. This helps explain why they can appear artificial when framed as learning problems.

The evidence classification of~\citet{lin2026quantum} provides yet another coordinate. It grades application classes from Level~I, comprising highly structured problems with provably expensive classical cost, to Level~IV, comprising problems with strong classical baselines or limitations arising from input and output costs. This classification does not determine the mechanism profile. In particular, their use of ``provably expensive'' includes hardness relative to the best-known classical algorithm, whereas our criterion for establishing interference as necessary requires a proved bound in a specified access model. A Level~I rating therefore does not, by itself, establish that interference is necessary. Practical filters such as verifiability and super-quadratic speedup~\citep{babbush2026grand} likewise assess the evidence or quality of an advantage rather than identifying its mechanism.

Finally, algorithm catalogues such as the quantum algorithm zoo~\citep{jordan_zoo} organize methods by technique, while speedup taxonomies classify advantages by magnitude, such as polynomial, superpolynomial, or exponential. These are useful organizational tools, but neither directly answers the question that motivates our taxonomy: which counterfactual restriction would make a particular claimed advantage disappear? Our contribution is to make that question explicit, apply it across heterogeneous claims, and record mechanism presence separately from necessity and evidence strength.


\section{Reporting hygiene}
\label{sec:hygiene}

The audit suggests four reporting checks. None is specific to quantum computing, and each identified at least one problematic claim.

First, state the adversary. A lower bound applies to a defined class of strategies, so name that class, whether all classical algorithms in a query model, memoryless quantum learners, measure-and-prepare channels, bounded-depth circuits, statistical queries, or the particular methods tested. Also state whether the bound is proved in the paper, imported from another result, or conjectured. A quantum optimality bound is not a classical lower bound. ``Measure-first'', ``incoherent'', and ``single-copy'' describe different adversaries~\citep{gyurik2023limitations,huang2021information,jerbi2023incoherent}. The error measure matters as well: average-case and worst-case prediction can yield different conclusions about a separation~\citep{huang2021information}.

Second, distinguish convergence failure from hardness. A tensor-network simulation that fails to converge, or a Pauli-path truncation whose result remains dependent on the truncation level, establishes a limitation of current methods at the tested sizes, not a general classical lower bound. Such results belong in the evidence register as benchmarks, as the papers reporting them carefully acknowledge. Noise can change the asymptotic picture in the opposite direction. At a constant depolarizing rate per gate, random circuit sampling in the anti-concentrated regime is classically samplable in polynomial time~\citep{aharonov2022polynomial}. Although the algorithm is impractical, with an exponent that grows as the inverse noise rate, and its authors do not claim to address finite-size experiments, this result limits asymptotic claims of interference-based advantage for noisy sampling. Such claims require either a finite-size benchmark, reported explicitly as empirical evidence, or error suppression that improves as circuit size grows. The same caution applies when hardness is argued relative to the best-known algorithm: identify the rung of the hardness ladder relevant to the operating point, since some proposed operating points carry no complexity-theoretic guarantee even when other rungs have rigorous support.

Third, check the conditioning variables. Ask whether the claimed separation depends on a particularly favourable configuration, such as a specific input model, a fitted exponent extrapolated beyond its validated range, or a parameter fixed at its optimal value. Then assess whether the quantity distinguishing the quantum and classical methods remains meaningful under a realistic noise level and shot budget.

Finally, report no more precision than the inputs support. A speedup ratio quoted to two significant figures can imply unjustified precision when the variance of runtimes across shot noise, calibration drift, and baseline tuning has not been reported. Averaging can reduce a measured spread, but selecting a point estimate from that spread does not eliminate the underlying uncertainty.


\section{Limits and open questions}
\label{sec:limits}

The mechanism profile classifies proposed sources of advantage, not their validity. It records plausible mechanisms and the counterfactuals under which their contributions would disappear, but does not establish that a claimed separation exists. That is the role of the evidence register. For each mechanism, presence and necessity must be reported separately, and a profile may retain several labels.

The audit is limited in both size and scope. It covers fifty papers collected for other purposes, initially labelled by one reader and then revisited using the same procedure. The proportions in Table~\ref{tab:audit} characterize this sample and the corrections made during screening, not the literature as a whole. The audit instead illustrates which reporting errors the procedure can catch, including errors introduced when claims are screened from summaries rather than examined in their full context.

The distinction between coherent-memory and coherent-output mechanisms emerged when a witness involved an adversary not captured by the original test. Other separations, including those against bounded classical memory or bounded circuit depth, are currently recorded in terms of their resource-restricted adversaries. If enough such results accumulate, space may warrant its own mechanism label. Noise is another relevant axis. Without error correction, constant-depth interference can survive constant local noise in query models: a noisy device still solves Bernstein--Vazirani with $O(\log n)$ queries~\citep{chen2022nisq}. The separation between shallow quantum and classical circuits also survives local stochastic noise below a constant threshold, but there error correction is built in: the circuit acts on a surface code prepared single-shot, and the decoding is absorbed into the relation problem~\citep{bravyi2020noisy}. Deeper interference-based protocols can lose their advantage~\citep{chen2022nisq,aharonov2022polynomial}, and memory advantages can be damped exponentially~\citep{chen2022nisq}. These considerations suggest that the present labels may need to be refined as the range of audited results grows.

A further possible class concerns advantages attributed to the spectral geometry of operators on graphs or complexes, including curvature, Hodge Laplacians, and collective modes. Such proposals do not fit neatly into structural compression, within-register interference, or cross-copy coherence. However, none of the cases considered here establishes a proved separation, and some may ultimately reduce to structural conditions on the input model. We therefore identify geometric mechanisms as a possibility for future investigation, not as an established category in the current taxonomy.

Finally, profiles should remain revisable. Each mechanism entry should specify the counterfactual that could change its status: a dequantization for a proposed structural contribution, a better classical algorithm or a proved bound for an interference candidate, or efficient copy-wise post-processing for a provisional memory contribution. The profile should make explicit both what evidence would change the classification and whether the current evidence establishes the mechanism as necessary.

\bibliographystyle{plainnat}
\bibliography{refs}

\end{document}